\documentclass[prd,superscriptsize,superscriptaddress,onecolumn,longbibliography,nofootinbib,notitlepage]{revtex4-1}
\usepackage[utf8]{inputenc} 

\usepackage{graphicx,xcolor,overpic,mathtools}
\usepackage{amsthm,amsmath,amssymb}
\usepackage[colorlinks]{hyperref}
\definecolor{coolblack}{rgb}{0.0, 0.18, 0.39}
\hypersetup{
    colorlinks = true,
    citecolor = {coolblack},  
    urlcolor = {blue}
   }
\PassOptionsToPackage{normalem}{ulem}
\usepackage{ulem} 
\usepackage{sidenotes}
\usepackage{bm}
\usepackage{url}
\usepackage{physics}
\usepackage{xfrac}
\usepackage[makeroom]{cancel}
\usepackage{tensor}
\usepackage{mathrsfs}
\usepackage{slashed}
\usepackage{tikz}
\usepackage[mode=buildnew]{standalone}
\usepackage{caption,subcaption}

\newcommand{\fpi}{f_{\pi}}
\newcommand{\mpi}{m_{\pi}}

\newcommand{\mcal}{\mathcal{M}}
\DeclareMathOperator{\MeV}{MeV}

\DeclareMathOperator{\fm}{fm}

\newcommand{\comment}[1]{}

\NewDocumentCommand{\evat}{sO{\bigg}mm}{%
  \IfBooleanTF{#1}
   {\mleft. #3 \mright|_{#4}}
   {#3#2|_{#4}}%
}
\usepackage{enumerate}
\usepackage{cleveref}
\usepackage{todonotes}

\usepackage{stackengine}
\stackMath

\begin{document}
\title[
]{The charge density and neutron skin thickness of Skyrmions}

\author{Alberto García Martín-Caro}
\email{agmcaro@gmail.com}
\affiliation{Departamento de F\'isica de Part\'iculas, Universidad de Santiago de Compostela and \\Instituto
Galego de F\'isica de Altas Enerxias (IGFAE) E-15782 Santiago de Compostela, Spain}
\affiliation{Department of Physics, University of the Basque Country UPV/EHU, Bilbao, Spain}
\author{Chris Halcrow}
\email{chalcrow@kth.se}
\affiliation{Department of Physics, KTH-Royal Institute of Technology, 10691 Stockholm, Sweden}

\date[ Date: ]{\today}
\begin{abstract} 
Motivated by recent parity violating electron scattering experiments, we compute the Neutron Skin Thickness (NST) of nuclei modelled as (quantized) Skyrmions, the topological solitons of the Skyrme model. We show how in a certain approximation, the result for the NST is oblivious to the fine details of the (generally very complicated) quantum state of the soliton and only depends on the total baryon number and the isospin number. Moreover, in the leading order, the linear dependence on the asymmetry parameter is recovered, as expected both from experimental data and other models of nuclei such as the liquid drop model.
\end{abstract}
\maketitle
\comment{
\begin{minipage}{\textwidth}
\tableofcontents
\end{minipage}
}
\section{Introduction}

Some properties of nuclei, such as the energies of their quantum ground states and their charge radii, have long been experimentally measured with outstanding accuracy. In particular, electromagnetic Form Factors (FFs), which describe how the nucleon target reacts in an elastic scattering with an electrically charged (usually leptonic) probe and contain information about the internal distribution of charge and magnetization within nuclei, have been used to determine the nuclear structure for more than fifty years \cite{RevModPhys.28.214}. Electromagnetic FFs of nucleons and nuclei have been measured over the last decades in elastic electron scattering experiments with an increasing precision \cite{DeJager:1974liz,Perdrisat:2006hj}, reaching an impressive level recently at JLab \cite{Punjabi:2015bba, Camsonne:2014, Arrington:2023hht} even for large momentum transfer. 

Since they were first proposed, it has been customary to interpret {FFs} as the Fourier transforms of local spatial distributions. However, relativistic wave functions are frame-dependent, and hence such an interpretation is often restricted to the Breit frame \cite{Sachs}.
Recently, the generalization of this interpretation to an arbitrary, relativistic reference frame was proposed in \cite{PhysRevLett.125.232002} by introducing an appropriate kinematic factor in the Fourier integral (which also resolves the apparent contradiction between the same magnitude obtained in the Breit and infinite momentum frames \cite{Miller:2007uy}).

The issue of whether it makes sense to define a localized charge density for quantum objects such as nucleons has also been subject to some debate \cite{Jaffe:2020ebz}. This ambiguity was recently resolved in \cite{Epelbaum:2022fjc}, where a redefinition of the charge density in terms of a new integral involving the electric form factor was shown to not depend on the particular wavefunction, as long as one can assume its spherical symmetry, i.e. in the rest frame of the system. 

Therefore, an accurate description of the local charge densities of nucleons, and especially of the proton, has become one of the main goals of theoretical nuclear physics. The description is not only essential to the knowledge of strong forces in the non-perturbative regime but also to the understanding of other precision observables in quantum electrodynamics \cite{Gao:2021sml} and other fundamental interactions, including gravity \cite{Polyakov:2018zvc}. However, compared to electromagnetic charge densities, other properties of nuclei, such as the neutron spatial distribution, are even less known experimentally. Indeed, while the proton distribution is explored through electromagnetic probes (charged leptons), whose interactions are well understood from the theory of quantum electrodynamics, the distribution of neutrons has been explored mainly through hadronic probes such as pions \cite{KRASZNAHORKAY2004224} or heavy ions \cite{Xu:2022ikx,PhysRevLett.125.222301}. These are described by quantum chromodynamics, which is subject to large uncertainties in the low energy regime due to its non-perturbative nature.  

Both kinds of charge densities are a valuable source of information about the internal structure of nuclei and the strong force. The distribution of charges inside a nucleus can reveal information about the isospin-asymmetric nuclear forces \cite{Mondal:2016bls}.  It turns out that these are crucial to the understanding of matter inside compact stars, and play a relevant role in the nuclear matter Equation of State \cite{Lattimer:2023rpe}.

A related observable is the \emph{neutron skin thickness} (NST), defined as the difference between the mean square radius of neutrons and that of protons, i.e.
\begin{equation}
    \Delta R_{np}(^BX)=R_n-R_p,\qquad R_{n}=\sqrt{\frac{\int \rho_n r^2d^3x}{B-Z}},\quad R_{p}=\sqrt{\frac{\int \rho_p r^2d^3x}{Z}},
\end{equation}
where $R_{n\,(p)}$ represents the root mean square (rms) radius of the point neutron (proton) distributions, $\rho_{n(p)}$.
The NST has been discussed for a long time in the nuclear physics literature (for a recent review on the topic see eg. \cite{Sammarruca:2023mxp}). Traditionally,  theoretical discussions on the NST have not been enough to determine its structure in a model-independent fashion, and the difficulty of observing the neutron distribution experimentally translated into a poor understanding of the quantity.

However, there has recently been an increased interest in the NST since parity-violating electron scattering (PVES) experiments aiming to observe the neutron distribution of $^{208}\rm Pb$ \cite{PREX} and $^{48}\rm Ca$ \cite{CREX} have been successfully performed. In such experiments, an electron is scattered off a nucleus, and the tiny asymmetry between right and left-handed scattered leptons due to parity-non conservation of the weak force is measured. 
The measurement of the parity violating asymmetry in polarized scattering allows one to determine the nuclear weak charge form factor, interpreted as the Fourier transform of the weak charge density for a given momentum transfer, that is also strongly correlated with the neutron distribution (since the weak neutral coupling of protons is much smaller than that of neutrons at low momentum transfer, i.e $Q^p_W = 1 - 4\sin^2\theta_W\approx 0.075\ll Q^n_W = -1$).
Although such measurements avoid strong interaction uncertainties by using leptonic probes, they are technically very challenging as they are based on a very small effect, of order about a part per million \cite{Horowitz:1999fk,Roca-Maza:2011qcr}.

Additionally, other leptonic probes can be used to study the weak charge distribution on nuclei. In particular, experiments involving neutrinos, whose interaction with nuclear matter is purely mediated by neutral weak bosons, have been proposed in order to measure the neutron density distribution of nuclei via the so-called Coherent Elastic neutrino-Nucleus Scattering  (CEvNS) \cite{Freedman:1973yd,Patton:2012jr,Payne:2019wvy}. As opposed to the PVES, in which the scattered electrons are measured, the outgoing neutrinos are impossible to detect.
Instead, what is measured is the very low kinetic energy of the recoiling nucleus. Due to the difficulty of such measurement, its experimental realization had to wait four decades, until the COHERENT collaboration was able to observe CEvNS \cite{COHERENT:2017ipa} (see also \cite{COHERENT:2020iec}).

Inspired by this recent experimental progress, we will use the Skyrme model to compute the neutron skin thickness of nuclei with arbitrary size and isospin, in the limit of small isospin asymmetry (i.e. small ratio of protons versus neutrons). In the Skyrme model approach, nuclei are semi-classical objects: one starts with a classical, localized topological soliton solution (called Skyrmions), and first quantizes its zero-modes to find the allowed quantum states representing physical nuclear states. This is the so-called \emph{rigid rotor quantization}. Using this approximation, the calculation of the charge densities and radius for the lightest nuclei, such as the $B=2$ (deuteron) and $B=3$ (triton and helium-3) solutions was performed more than thirty years ago \cite{Braaten:1988bn,Carson:1991fu}.
Although the Skyrme model provides a systematic way of obtaining quantum states and their corresponding charge densities for larger nuclei, in practice  the computations become very involved for sufficiently large baryon number, hence there have been very little progress in the study of local charge densities for large nuclei from the Skyrme model approach. Some limited investigations have been performed involving magnitudes that do not require quantization, such as the charge densities of isospin-zero states \cite{Karliner:2015qoa} or gravitational form factors \cite{GarciaMartin-Caro:2023klo}.
However, large, neutron-rich nuclei are precisely the most interesting to experimentally determine properties involving isospin asymmetric pat of nuclear interactions, such as the NST.

In this paper, we use the rigid rotor approximation and find an expression for the NST that only depends on simple properties of the classical solutions, (and not on the complicated details of the quantum states) which are easily computed numerically. This result can be derived directly without doing any intermediate steps involving the associated form factors, therefore avoiding the above-mentioned problems with their interpretation as Fourier transforms of real densities.

\section{The Skyrme model: Skyrmions as nuclei}

The Skyrme model that we consider is given by the following Lagrangian density,
\begin{align}
    \mathcal{L}_{\rm SK}=
    -\frac{\fpi^2}{16\hbar}\Tr\{L_{\mu}L^{\mu}\} + \frac{\hbar}{32e^2}\Tr\{[L_{\mu}, L_{\nu}]^2\} + \frac{\mpi^2\fpi^2}{8\hbar^3}\Tr\{U - \bm{1}\},
    \label{Lagrangian}
\end{align}
where we use the metric convention $\eta^{\mu\nu}={\rm diag}(1,-1,-1,-1)$. $U\in SU(2)$ contains the fundamental mesonic degrees of freedom, parametrized as the coordinates of the $SU(2)$ group element
\begin{equation}
    U=\sigma \bm{1}+i\pi_a\tau_a\equiv i\phi^\alpha\bar\tau_\alpha,\quad \phi^\alpha =(\sigma,\bm{\pi}),\quad \Bar{\tau}_\alpha=(-i\bm{1},\bm{\tau}),
\end{equation}
with $\tau^a$ ($a=1,2,3$) being the Pauli matrices. 
The left-currents $L_{\mu} = U^{\dagger}\partial_\mu U$ are the components of the associated left-invariant Maurer-Cartan form, $\bm{1}$ is the $2\times2$ identity matrix, and $B^{\mu}$ is the topological current,
\begin{equation}
    B^{\mu} = \frac{\epsilon^{\mu\nu\rho\sigma}}{24\pi^2}\Tr\{L_{\nu}L_{\rho}L_{\sigma}\}.
\end{equation}
The integral of the zeroth element of the current is equal to the topological charge of the skyrmion
\begin{equation}
 B = \int   B^0(x) \, d^3x = \int \frac{\epsilon^{ijk}}{24\pi^2}\Tr\{L_{i}L_{j}L_{k}\}\, d^3 x \in \mathbb{Z} \, .
\end{equation}

We will consider static solutions of the Skyrme field and for numerical purposes, we adopt the usual Skyrme units of energy and length,
\begin{equation}
    E_s = \frac{ \fpi}{4e} , \quad x_s = \frac{2\hbar}{\fpi e},
\end{equation}
and use the dimensionless pion mass
\begin{equation}
    m = \frac{2 m_\pi   }{f_\pi e} \, .
\end{equation}
The static energy functional in these units becomes
\begin{equation}
    E[U] = \int d^3x \left[ -\frac{1}{2}\Tr\left\{L_iL_i\right\} - \frac{1}{16}\Tr\left\{\left[L_i,L_j\right]\left[L_i,L_j\right]\right\} + m^2 \Tr\{\bm{1} - U\} \right]  \, .
    \label{Energy_integral}
\end{equation}


\subsection{Quantization}
Nucleons and nuclei are described within the Skyrme model as classical solitonic configurations through the identification of the Skyrmion topological charge with the baryon number of nuclear states. However, other quantum numbers such as the spin and isospin of quantum nuclear states are not described at the classical level. Hence, a quantization of the Skyrmion field is needed to take into account the relevant quantum numbers. This is done in the semi-classical approach by promoting the zero modes of the soliton to dynamical degrees of freedom.

To do so, we introduce the rotational and iso-rotational degrees of freedom through a pair of time-dependent $SU(2)$ transformations of the classical (static) solitonic solution, 
representing the iso-rotation and the spatial rotation zero modes, respectively,\footnote{Skyrmions also present translational zero modes. However, these won't play any role in this work and we omit them for simplicity.}
\begin{equation} 
    U(t,\bm{x})= A(t) U_0(R_B(t)(\bm{x}))A^\dagger (t) \, ,
    \label{transformiso}
\end{equation}
where $R^{ij}_B=\tfrac{1}{2}\Tr{\tau^i B\tau^jB^\dagger }\in SO(3)$ is the corresponding rotation matrix in space.  $A(t),B(t)\in SU(2)$ 
form the \emph{collective coordinates} of the soliton. We note that, for spherically symmetric solutions,  rotation in coordinate space can be undone by that in isospin space, reducing the effective number of independent degrees of freedom \cite{Adkins:1983ya}. However, the present treatment is more general and can also be used for non-spherical solutions that we shall be mainly interested in.  
The semi-classical quantization of the Skyrmion then consists of substituting \eqref{transformiso} into the Skyrme Lagrangian \eqref{Lagrangian}, which yields the Lagrangian of an effective dynamical system in terms of the collective coordinates $\{A(t), B(t)\}$, and quantizing such a system via standard canonical methods. Performing the substitution yields the following Lagrangian
\begin{equation}
    L_{\rm col}=\int d^{3} x \mathcal{L}_{\rm SK}=-E[U_0]+\frac{1}{2} a_{i} U_{i j} a_{j}-a_{i} W_{i j} b_{j}+\frac{1}{2} b_{i} V_{i j} b_{j},
    \label{collectivelag}
\end{equation}
where 
$$
a_{j}=-i \operatorname{Tr} \tau_{j} A^{-1} \dot{A}, \quad b_{j}=i \operatorname{Tr} \tau_{j} \dot{B} B^{-1}
$$
are the angular velocities in isospace and physical space, respectively. We have also introduced the corresponding $3\times 3$ inertia tensors $U, W$ and $V$ defined by 
\begin{equation} \label{eq:MOI}
\Lambda = \begin{pmatrix} U & W \\ W^T & V \end{pmatrix}, \quad \Lambda_{ij} = -\int \frac{1}{2} \text{Tr}\left( G_i G_j \right) + \frac{1}{8}\text{Tr}\left( [L_k, G_i][L_k,G_j] \right)  d^3x \, ,
\end{equation}
where $G_i = iG_{ia}\tau_a$ is an $\mathfrak{su}(2)$ current equal to
\begin{equation} \label{eq:Gdef}
G_i = \begin{cases}  \tfrac{i}{2}U_0^\dagger[ \tau_i, U_0 ]\, , \quad& i=1,2,3 \\ \epsilon_{ilm}x_l L_m\, , \quad& i=4,5,6\,, \, . \end{cases} 
\end{equation}
For notational simplicity, here and below we do not distinguish upper ($x^{i},\tau^a$) from lower ($x_{i},\tau_a$) indices for three-dimensional vectors and tensors when we deal with purely three-dimensional expressions. 

The quantum operators of the angular velocities $a_{j}$ and $b_{j}$ correspond to the rotational and iso-rotational collective coordinates $A,B$ and their canonically conjugate momenta, the body-fixed isospin and spin angular momentum operators $K_{j}$ and $L_{j}$. These are obtained in terms of the angular velocities via the relations,
\begin{equation}
\begin{aligned}
K_{i} &=U_{i j} a_{j}-W_{i j} b_{j} \\
L_{i} &=-W_{j i} a_{j}+V_{i j} b_{j} \, .
\label{momentum-velocities}
\end{aligned} 
\end{equation}
These operators are related to the usual space-fixed isospin and spin angular momentum operators $I_{j}$ and $J_{j}$ via
\begin{equation}
I_{i}=-R_{i j}(A) K_{j}, \quad J_{i}=-R_{i j}\left(B\right)^{T} L_{j}
\label{body-fixed-relations}
\end{equation}
implying $\bm{I}^{2}=\mathbf{K}^{2}$ and $\bm{J}^{2}=\mathbf{L}^{2}$. The set of operators, $\bm{I}, \bm{J}, \bm{K}$ and $\bm{L},$ form an irreducible representation of the Lie algebra of $\mathcal{O}_{I, K} \otimes \mathcal{O}_{L_{n} J},$ the symmetry group of two rigid rotators, and obey the commutation relations
\begin{equation}
{\left[I_{i}, I_{j}\right]=i \epsilon_{i j k} I_{k},}\quad  {\left[K_{i}, K_{j}\right]=i \epsilon_{i j k} K_{k}} ,\quad
{\left[J_{i}, J_{j}\right]=i \epsilon_{i j k} J_{k 1}} ,\quad {\left[L_{i}, L_{j}\right]=i \epsilon_{i, j k} L_{k}}.
\end{equation}
From these, we may produce a complete set of commuting observables to define an eigenstate basis of $\mathcal{H}_G$. We construct such a basis with states of the form
\begin{equation}
    \ket{G}=\ket{i,i_3,k_3}\otimes\ket{j,j_3,l_3}\equiv\ket{i,i_3,k_3;j,j_3,l_3}
    \label{basisG}
\end{equation}
where $j$ and $j_3$ correspond to the eigenvalues of the corresponding total angular momentum and the third component of angular momentum operators, where $-i \leq i_{3}, k_{3} \leq i$ and $-j \leq j_{3}, l_{3} \leq j .$ In particular, the subspace we will be mostly interested in is that of fixed $i, i_{3}, j, j_{3}$, labeled by the states $\ket{k_{3}, l_{3}},$ which is $(2 i+1)(2 j+1)-$dimensional.

\section{The electric charge density of Skyrmions}

The Gell-Mann-Nishijima formula tells us how to obtain the charge density of a Skyrmion field configuration,
\begin{equation}
    Z = \frac{1}{2}B + I_3 \hspace{2mm} \longrightarrow \hspace{2mm} \rho = \frac{1}{2}B^0 + \ev{I^0_3},
    \label{charge_dens}
\end{equation}
where $I^0_3$ is the time-like component of the third isospin Noether current, and the brackets represent the expectation value on the quantum state of the Skyrmion. An explicit expression for (the classical version of) this current can be obtained from Noether's theorem, given that an infinitesimal isospin transformation
\begin{equation}
    U\rightarrow U'=U+\epsilon^k\delta U_k,\qquad \delta U_k=\frac{i}{2}[\tau^k,U],
\end{equation}
generates the Noether current 
\begin{equation}
    I^\mu_{k}=-\frac{1}{24\pi^2}\big[ \Tr{L^{\mu} G_{k}}+\Tr{\left[L_{\nu}, L^{\mu}\right]\left[L^{\nu}, G_{k}\right]}\big] ,\label{NoetherVCurr}
    \end{equation}
where $G_k$ is the $\mathfrak{su}(2)$ current defined in \cref{eq:Gdef} ($k=1,2,3$).

The quantum version of $I_3^0$ can be obtained in rigid-rotor quantization by substituting the dynamical ansatz \eqref{transformiso} into \eqref{NoetherVCurr}. One then replaces the classical variable $a_i$ with the corresponding quantum operators and applies Weyl ordering to the products of two or more non-commuting operators \cite{Carson:1991fu}. In the case of the angular velocity, we will use the quantum (body-fixed) angular momentum operators i.e. we need to invert \eqref{momentum-velocities}. This simplifies significantly if we assume that the moment of inertia tensors satisfy the following sphericity condition:
\begin{equation}
    U_{ij}=u\delta_{ij},\quad V_{ij}=v\delta_{ij},\quad  W_{ij}=w\delta_{ij}.
    \label{eq:sphericity}
\end{equation}
i.e, the tensors defined in \cref{eq:MOI}  are proportional to the identity. This is true for the $B=1,3$ skyrmions and can be a good approximation for large skyrmions. Indeed, \cref{eq:sphericity} deals with quantities defined on the internal isospin space, and can be a good approximation even for highly non-spherical solutions in real space, as we show in the next section.

The expectation value of $I_3^0$ in the isospin state $\ket{\psi}$ is then given by
\begin{align}
    \ev{I_3^0}=&-\frac{1}{(uv-w^2) }\bra{\psi}\qty[R_{3a}u_{ab}(vK^b+wL^b)-R_{3a}w_{ab}(wK^b+uL^b)]_{\rm Weyl}\ket{\psi}=\nonumber\\
    =&-\frac{1}{2}\bra{\psi}\qty[\frac{(u-v)u_{ab}+(u-w)w_{ab}}{(uv-w^2) }[R_{3a}K_{b}]_++\frac{wu_{ab}+uw_{ab}}{(uv-w^2) }[R_{3a},M_{b}]_+]\ket{\psi}
\label{eq:I3ev}
\end{align}
where we have used the relation between the body-fixed and space-fixed angular momenta \eqref{body-fixed-relations}, and defined the grand spin operator $M_i=K_i+L_i$.
The non-trivial part of the calculation is to compute the matrix elements
\begin{equation}
    \mathcal{M}_{ab}(\psi)=\frac{1}{2}\bra{\psi}[R_{3a},K_{b}]_+\ket{\psi},\qquad \tilde{\mathcal{M}}_{ab}(\psi)=\frac{1}{2}\bra{\psi}[R_{3a},M_{b}]_+\ket{\psi} \, .
\end{equation}
If this can be done, one can obtain the charge density of the skyrmion in any given quantum state $\ket{\psi}$. 

In the case of nucleons, spherical symmetry implies $u=v=w$, so \cref{eq:I3ev} is undefined. To get rid of the apparent divergence we can rewrite the denominator as 
\begin{equation}
   \lim\limits_{v,w \to u} uv-w^2=(u+u)(u-u)
\end{equation}
and the term $(u-u)$ cancels with the same term in the denominator, once we take into account the fact that nucleons are grand spin singlets, i.e. satisfy $\vec{L}\ket{\psi}=-\vec{K}\ket{\psi}$.
Therefore, for nucleon states $\ket{\psi}=\ket{\frac{1}{2},i_3,j_3}$ \cite{BraatenB1EW}, the expectation of $I^0_3$ is
\begin{equation}
\ev{I^0_3}{\psi}=-\frac{u_{ij}}{2u}\bra{\tfrac{1}{2},i_3,j_3}[R_{3j}K_{i}]_+\ket{\tfrac{1}{2},i_3,j_3}=-\frac{u_{ij}}{6u}\delta^{ij}\ev{\tau^3}{i_3}\braket{j_3}{j_3} \, .
\end{equation}
Thus the charge density of quantized $B=1$ Skyrmions in the ground state is given by
\begin{equation}
    \rho(\vec{x};i_3) =\frac{1}{2}B^0(r)-i_3\frac{\delta^{ij}u_{ij}(r)}{3u}
\end{equation}
which preserves spherical symmetry.

\begin{figure}
    \centering
    \includegraphics[scale=0.6]{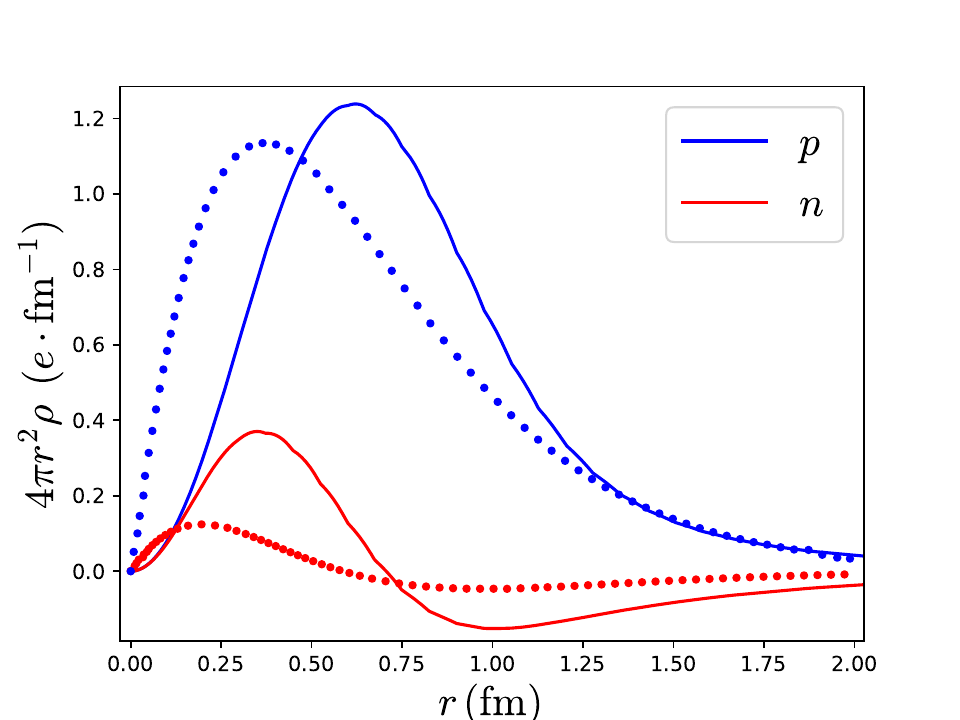}
    \caption{Radial profile of the electric charge density of nucleons}
    \label{fig:chargedensB1}
\end{figure}

In \cref{fig:chargedensB1} we plot the radial profile of the charge densities for protons and neutrons as predicted by the Skyrme model. The dotted line data is calculated from the nucleon form factors, using an improved definition of the charge density \cite{Epelbaum:2022fjc}.

This method can be generalized to certain higher baryon charges. The $3-$skyrmion also obeys the sphericity condition \eqref{eq:sphericity}. The ground state of its physical Hilbert space is the spin $\tfrac{1}{2}$, $( ^3{\rm H}, ^3{\rm He})$ isospin doublet \cite{Manko:2007pr, Carson:1990yv},
\begin{equation}
   \ket{\psi}\equiv \ket{^3{\rm H}/ ^3{\rm He},s}=\ket{\tfrac{1}{2},s,\tfrac{1}{2}}\ket{\tfrac{1}{2},\sfrac{+}{-}\tfrac{1}{2},\tfrac{-1}{2}}-\ket{\tfrac{1}{2},s,\tfrac{-1}{2}}\ket{\tfrac{1}{2},\sfrac{+}{-}\tfrac{1}{2},\tfrac{1}{2}}.
    \label{eq:Tritiumstates}
\end{equation}
The states are grand spin singlet, so $\Tilde{M}_{ij}$ will vanish, and we have
\begin{align}
    \bra{\psi}I_3^0(\vec{x})\ket{\psi}&=-\mcal_{ab}(\psi)\frac{(v-w)u_{ab}(\vec{x})+(u-w)w_{ab}(\vec{x})}{(uv-w^2) },\\[2mm]
    \rho(\vec{x},\psi) =\frac{1}{2}B_0(\vec{x})+\bra{\psi}I_3^0(\vec{x})\ket{\psi}&=\frac{1}{2}B_0(\vec{x})-i_3\delta_{ab}\frac{(v-w)u_{ab}(\vec{x})+(u-w)w_{ab}(\vec{x})}{3(uv-w^2) }.
\end{align}

\begin{figure}
    \hspace*{-2.5cm}
    \includegraphics[scale=0.25]{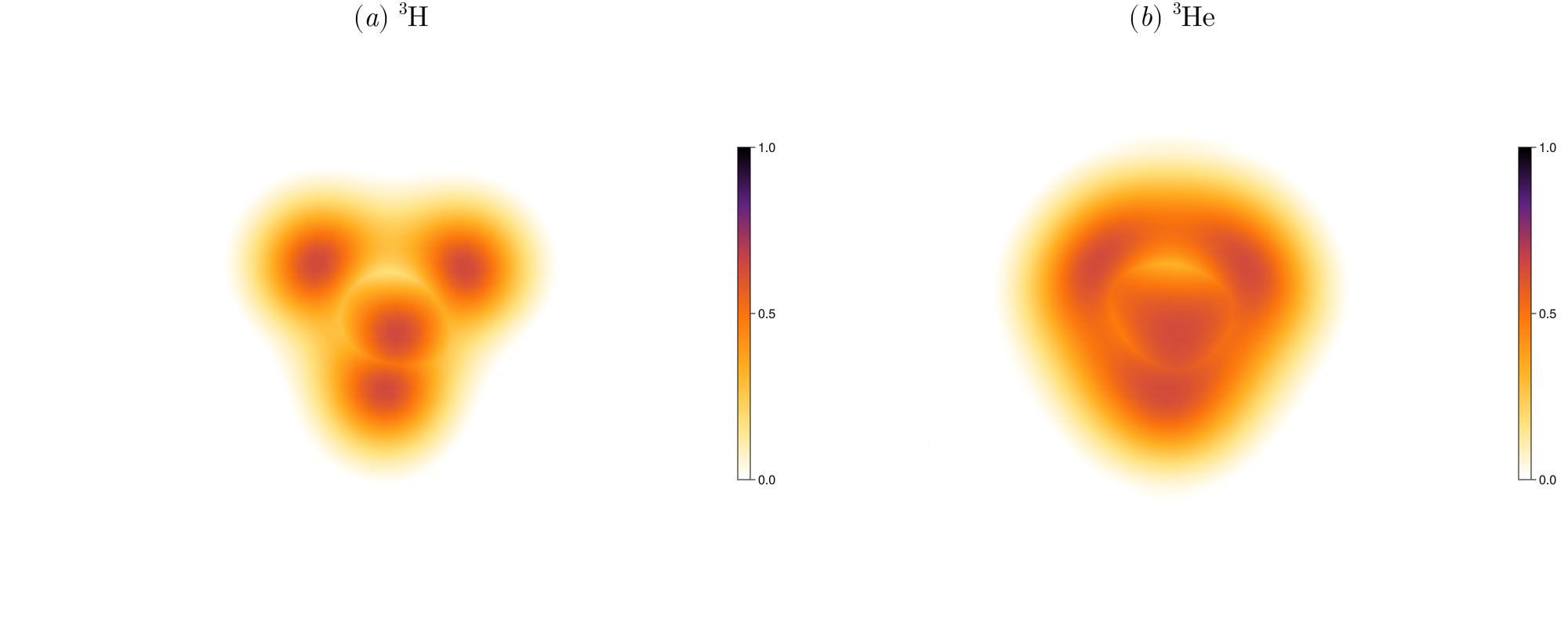}
    \caption{Charge densities of tritium and helium-3 ground states (in units of the electron charge)}
    \label{fig:HeH3densities}
\end{figure}

The charge densities of $^3 {\rm H}$ and $^3 {\rm He}$ ground states in the Skyrme model are not spherically symmetric, but preserve the tetrahedral symmetry of the classical solution (as required by the {FR} constraints). We plot a two-dimensional projection of these densities in \cref{fig:HeH3densities}.

\section{Neutron skin thickness of Skyrmions}

As we have seen, nuclei are described as classical topological solitons in the Skyrme model. By quantizing the skyrmions, we can compute an associated charge density for each quantum state. This is identified with an effective ``proton density''. Following a similar argument, the neutron density is given by 
\begin{equation}
    \rho_n=B_0-\rho_p=\frac{1}{2}B_0-\ev{I_3^0}
\end{equation}
so we can, in principle, compute the neutron skin thickness, $\Delta R_{np}$, of a given nucleus once we know its associated quantum state in the Skyrme model. It is given by
\begin{equation}
    \Delta R_{np} = R_n - R_p
\end{equation}
where
\begin{align}
R_n = \frac{1}{B-Z} \int \frac{1}{2}B_0(x)r^2 - \ev{I_3^0}r^2 \, d^3x \\
R_p = \frac{1}{Z} \int \frac{1}{2}B_0(x)r^2 + \ev{I_3^0}r^2 \, d^3x \, .
\end{align}

\subsection{Example: the 3-skyrmion}

As an example, consider the $B=3$ ground state $\ket{\psi}$ \eqref{eq:Tritiumstates}, which corresponds to the $(^3{\rm H}, ^3{\rm He})$ isospin doublet. We again use the fact that the state is a grand spin singlet, so $\Tilde{M}_{ij}$ will vanish, and so the proton radius $R_p$ is given by
\begin{align}
    R_{p}^2(\psi)&=\frac{1}{Z}\int\qty[\frac{1}{2}B_0r^2+\ev{I_3^0} r^2]d^3x\\
    &=\frac{1}{Z}\qty[\frac{1}{2}\int B_0r^2d^3x-\frac{M_{ij}(\psi)}{(uv-w^2) }\qty[(v-w)u_{ij}^{(2)}+(u-w)w_{ij}^{(2)}]],\notag
\end{align}
where we have defined the second moment of the inertia tensors
\begin{equation}
   u_{ij}^{(2)}\doteq \int u_{ij}(x)r^2d^3x,\qquad w_{ij}^{(2)}\doteq \int w_{ij}(x)r^2d^3x.
\end{equation}
Further, by the tetrahedral symmetry of the $B=3$
\begin{equation}
    u_{ij}^{(2)}=u^{(2)}\delta_{ij},\qquad  w_{ij}^{(2)}=w^{(2)}\delta_{ij} \, ,
\end{equation}
which allows us to use the identities $M_{ii}=\bra{\psi}R_{3i}K_i\ket{\psi}=-\bra{\psi}I_3\ket{\psi}=-i_3$, which simplifies the calculation significantly. The proton density rms radius is
\begin{equation}
    R_{p}^2(\psi)=\frac{1}{Z}\qty[\frac{1}{2}B^{(2)}+\frac{(v-w)u^{(2)}+(u-w)w^{(2)}}{(uv-w^2)}i_3],
    \label{Rp}
\end{equation}
where $B^{(2)} = B r_{rms}^2$, with $r_{rms}^2$ the rms matter radius.

Similarly, we can calculate the neutron density rms radius. It is
\begin{equation}
        R_{n}^2(\psi)=\frac{1}{3-Z}\qty[\frac{1}{2}B^{(2)}-\frac{(v-w)u^{(2)}+(u-w)w^{(2)}}{(uv-w^2) }i_3].
        \label{Rn}
\end{equation}
Overall, we have
\begin{align}
    R_{p}(^3{\rm He})=\sqrt{\frac{1}{4}\qty[B^{(2)}+\frac{(v-w)u^{(2)}+(u-w)w^{(2)}}{(uv-w^2) }]}=R_{n}(^3{\rm H}),\\
    R_{n}(^3{\rm He})=\sqrt{\frac{1}{2}\qty[B^{(2)}-\frac{(v-w)u^{(2)}+(u-w)w^{(2)}}{(uv-w^2) }]}=R_{p}(^3{\rm H}).
\end{align}

To find the neutron skin thickness of $^3$H, we must calculate $u, w, v, u^{(2)}$ and $w^{(2)}$. The final two quantities are quite difficult to calculate numerically. Their density scales as e.g. $ u_{ij} r^2$. If there is an $O(\epsilon)$ error in the Skyrme field, this error is also present in the $u_{ij}$ density and the $r^2$ further compounds the error. To gain confidence in our numerical methods, we calculated the quantities in the rational map approximation, using the identities in \cite{Manko:2007pr}. To compare, fix the parameters $\fpi= 131.3 \MeV$, $e=4.628$, and the physical value of the pion mass. The rational map skyrmion has $u_\text{RM}^{(2)} = 770.55 \text{ MeV fm}^4$ and $w_\text{RM}^{(2)} = -466.36 \text{ MeV fm}^4$ while our full field numerics give $u^{(2)} = 722.57 \text{ MeV fm}^4$ and $w^{(2)} = -424.43 \text{ MeV fm}^4$. Noting that the rational map skyrmions are approximations, the near agreement between calculations gives us confidence in our numerical method.

Overall, the neutron skin thickness of $^3$H in the Skyrme model is
\begin{equation}
    \Delta R_{np}(^3{\rm H})= -R_{np}(^3{\rm He})\approx 0.23 {\rm \,fm}.
\end{equation}
The measured charge radii of $^3 {\rm He}$ and $^3 {\rm H}$ are $1.959\pm 0.03\fm$ and $1.755\pm 0.086\fm$, respectively \cite{SICK2001245}.
Assuming perfect isospin symmetry, their difference measures the {NST} since
\begin{equation} 
\Delta R_{np}(^3{\rm H}) = R_n(^3{\rm H}) - R_p(^3{\rm H}) = R_p(^3{\rm He}) - R_p(^3{\rm H}) \, ,
\end{equation}
giving the experimental result $\Delta R_{np}(^3{\rm H})\eval_{\rm exp}\approx 0.2\fm$. For our choice of parameters, the predicted difference in charge radii of the $^3 {\rm H}-^3 {\rm He}$ isodoublet comes out very close to the experimental value.

\subsection{Large nuclei}

\begin{figure}
    \centering
    \includegraphics[width=0.5\linewidth]{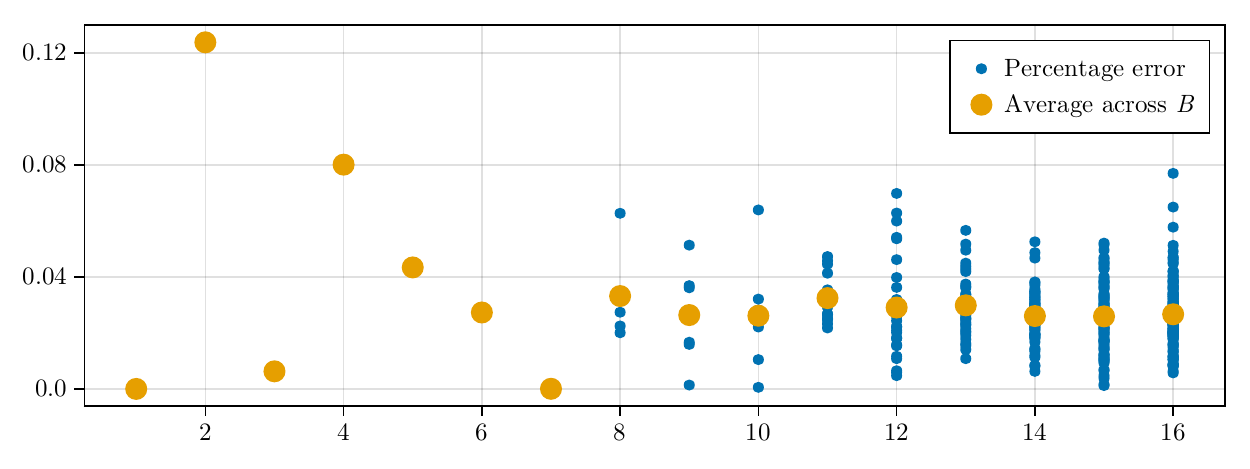}
    \caption{The percentage error in the iso-sphericity approximation $u_{ab} = u \delta_{ab}$ as a function of $B$, for the 409 skyrmion solutions found in \cite{Gudnason:2022jkn}.}
    \label{fig:errors}
\end{figure}

We now attempt to compute the {NST} for larger nuclei, without specific details of their quantization. The difficult part is to calculate the expectation value of $I_3^0$, given by
\begin{align} \label{eq:I3}
    \ev{I_3^0}= -\frac{1}{2}\bra{\psi}\qty[\frac{(u-v)u_{ab}+(u-w)w_{ab}}{(uv-w^2) }[R_{3a}K_{b}]_++\frac{wu_{ab}+uw_{ab}}{(uv-w^2) }[R_{3a},M_{b}]_+]\ket{\psi} \, .
\end{align}
It is possible to evaluate this expression for generic wavefunctions if we use three approximations:
\begin{equation} \label{eq:approximations}
u_{ab}(x) = u(x) \delta_{ab},  \quad |w_{ab}| \ll u , \quad u \ll v \, .
\end{equation}
To check the first approximation, which we call iso-sphericity, we have calculated its accuracy for the ``smorgasbord of skyrmions", a collection of 409 skyrmion solutions with $B=1-16$ \cite{Gudnason:2022jkn}. We plot the percentage error of the approximation in Fig. \ref{fig:errors}, defined as
\begin{equation}
    \text{err} = \frac{|| u_{ab} -  u \delta_{ab} || }{ || u_{ab} ||}, \quad \text{with} \quad || u ||^2 = \sum_{a,b} u_{ab}u_{ba} \, .
\end{equation}
The error is always below $8\%$ for $B>4$, and takes an average value of around $3\%$. Hence this seems to be a reasonable approximation to make. The mixed iso-term $w$ is generally small, and we will check the approximation numerically for the large skyrmions we create later in the paper. Generally, the spatial inertia tensor $V_{ij}$ grows as $B^2$ while $U_{ij}$ grows linearly with $B$. Hence we do expect the third approximation to be valid at large $B$.

We now apply the approximations, significantly simplifying \eqref{eq:I3} to
\begin{align}
    \ev{I_3^0}= -\frac{1}{2}\bra{\psi}\qty[ \frac{ u(x) }{u }[R_{3a}K_{a}]_+]\ket{\psi}  = \frac{i_3}{2}\frac{u(x)}{u} \, .
\end{align}
Most importantly, using the approximations means that we don't need to know the specific structure of the ground state wavefunction $\ket{\psi}$, only its isospin value.

We can now follow the same reasoning that leads to \cref{Rp,Rn}, and calculate the proton and neutron rms radii for an arbitrary Skyrmion of baryon charge $B$. They are
\begin{align}
    R_{p}^2(B,i_3)=\frac{1}{B+2i_3}\qty[B^{(2)}+2\frac{u^{(2)}}{u }i_3]= \frac{1}{1-\delta}\qty[r_{rms}^2-\frac{u^{(2)}}{u }\delta] , \label{Rp2}\\
    R_{n}^2(B,i_3)=\frac{1}{B-2i_3}\qty[B^{(2)}-2\frac{u^{(2)}}{u }i_3]= \frac{1}{1+\delta}\qty[r_{rms}^2+\frac{u^{(2)}}{u }\delta], \label{Rn2}
\end{align}
where 
\begin{equation}
    \delta=-\frac{2i_3}{B}=\frac{N-Z}{B}
\end{equation}
is the \emph{isospin asymmetry}, with $N$ and $Z$ the corresponding numbers of neutrons and protons in the nucleus. Given \cref{Rp2,Rn2}, we can get a rough estimate for the {NST} of \emph{any} nucleus from the Skyrme model, just by computing properties of the corresponding classical solution.

To understand the basic physics controlling the neutron skin, we can simplify further by assuming that $B\gg i_3$ (or, equivalently, $\delta\ll 1$. We insert this identity and remove a factor of $B/B$, to rearrange the radii as
\begin{equation}
R_p^2 \approx r_{rms}^2 -\delta\left( \frac{u^{(2)}}{u} - r_{rms}^2 \right) \implies R_p \approx r_{rms}\qty[1-\frac{\delta}{2}\qty(\frac{u^{(2)}}{r_{rms}^2 u} -  1)].
\end{equation}
Hence the {NST} is given by 
\begin{equation}
    \Delta R_{np} \approx \delta\, \left(\frac{u^{(2)}}{u r^2_{rms}} - 1  \right)r_{rms} \, .
\end{equation}
The quantity $u^{(2)}/u$ measures the root-mean-squared radius associated to the isospin density. Hence, the sign and magnitude of the neutron skin depends on the difference between the matter and isospin charge radii, $u^{(2)}/u-r^2_{rms}$.

Experimentally, the {NST} is measured to depend linearly on the isospin asymmetry, with a slope that seems to be independent of the baryon number in the range $16<B<238$ (see fig. 2 in \cite{Novario:2021low}) and parametrized by
\begin{equation}
    \Delta R_{np}= 1.32\,\delta-0.024 \pm 0.026\, {\rm fm}.
    \label{Exprel}
\end{equation}
In the next Section, we test if this linear behavior is reproduced for the Skyrme model. The relationship between $\Delta R_{np}$ and $\delta$ is expected to become nonlinear for small nuclei, in which few body interactions may dominate over bulk effects \cite{Novario:2021low}. It is precisely in this range of the baryon charge where the solutions of the Skyrme model have been extensively studied, including a recent exhaustive description of the landscape of all known local minima in the Skyrmion configuration space up to $B=16$ \cite{Gudnason:2022jkn}.

\section{Numerical Results}

Unfortunately, both the computational time required to find a minimum and the number of such local minima tend to grow very rapidly with the baryon number of the configurations, which makes it computationally very demanding to perform a systematic study of very large $(B\sim 100)$ Skyrmions. Nevertheless, we have been able to generate new classical solutions for Skyrmions with $B$ up to $256$ using the Skyrmions3D.jl package \cite{Halcrow_Skyrmions.jl}, which allows one to create and visualize 3D Skyrmions in the Skyrme model of nuclear matter, and compute their properties such as energy, baryon number, moments of inertia and the corresponding densities, etc. We have generated Skyrmions constructed from $B=4$ cubes, with $B=32, 48, 64, 108, 144, 180, 196, 240$ and  $256$. Each skyrmion is made on a cubic lattice with spacing $0.2$. The size of the lattice is adjusted depending on the size of the skyrmion. The largest lattice used was $160^3$ for the $B=256$ skyrmion; this is the largest-$B$ skyrmion ever created. We used Neumann boundary conditions and an arrested Newton Flow \cite{Battye:2001qn}. The flow is stopped when the maximum absolute value of the variation at \emph{any} lattice point is less than $0.01$.

\begin{figure}
    \centering
    \includegraphics[width=0.9\linewidth]{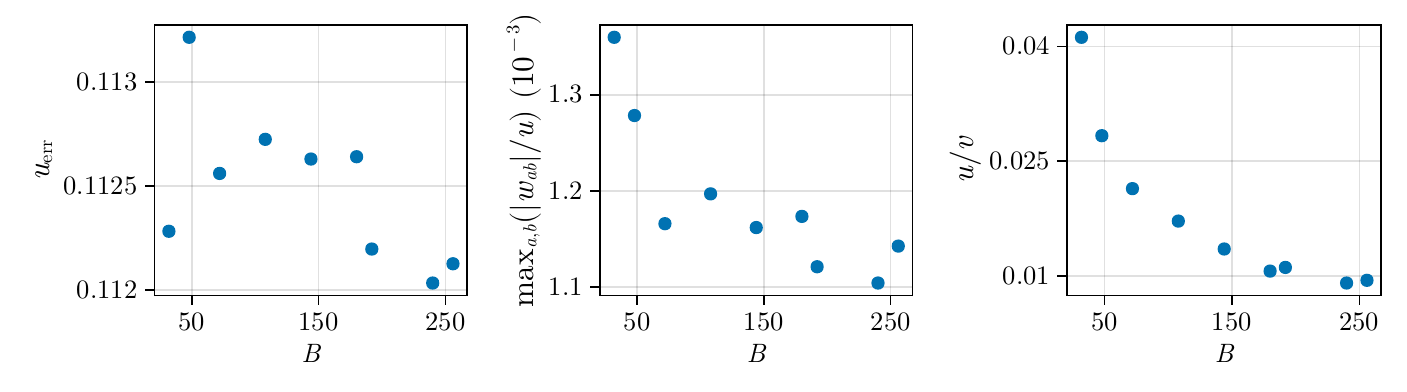}
    \caption{The percentage error in the iso-sphericity approximation $u_{ab} = u \delta_{ab}$, the value $\text{max}_{a,b}( |w_{ab}|/u)$ and the value $u/v$ for the numerically generated large skyrmions.}
    \label{fig:approximations}
\end{figure}

We first test the three approximations, listed in \eqref{eq:approximations}, for our numerically generated skyrmions. The percentage error in the sphericity approximation $u_{ab} = u \delta_{ab}$, the value $\text{max}_{a,b}( |w_{ab}|/u)$ and the value $u/v$ are plotted in Figure \ref{fig:approximations}. All these should be much less than one for our method to be accurate. Overall, the results are reassuring. The least accurate assumption is the iso-sphericity approximation, which has a $\sim 10\%$ error.

\begin{figure}
    \centering
    \includegraphics[width=0.9\linewidth]{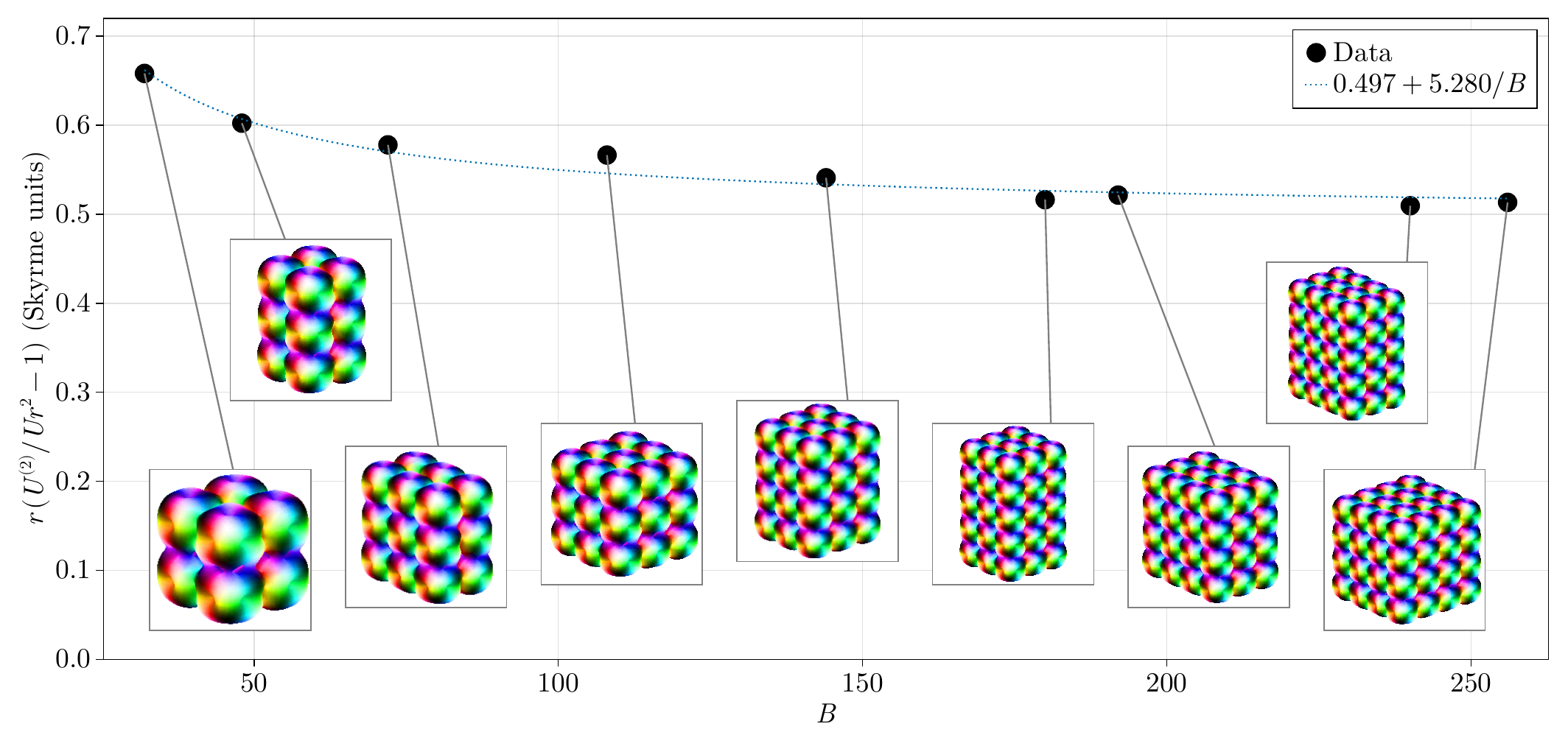}
    \caption{$r(U^{(2)}/Ur^2 - 1)$ in Skyrme units as a function of the baryon number, for $m=0.2$.}
    \label{fig:ruu}
\end{figure}

Using the numerically generated solutions, we can compute the {NST} of these large $B$ configurations and compare them to the experimental results. To do so, we calculate $\left(u^{(2)} / u r^2_{rms} - 1  \right)r_{rms} $ for the numerically generated Skyrmions. We plot these results, and energy densities of the generated skyrmion, in Fig. \ref{fig:ruu}. Remarkably, the value changes little over this large range of baryon numbers. From $B=64$ to $B=240$, the values change by around $10\%$. There is a clear decrease as $B$ increases, but the leading order behavior is constant. Hence the Skyrme model reproduces the linear relationship between $\Delta R_{np}$ and $\delta$. We then predict a correction, inversely proportional to $B$.

\begin{figure}
    \centering
    \includegraphics[width=0.7\linewidth]{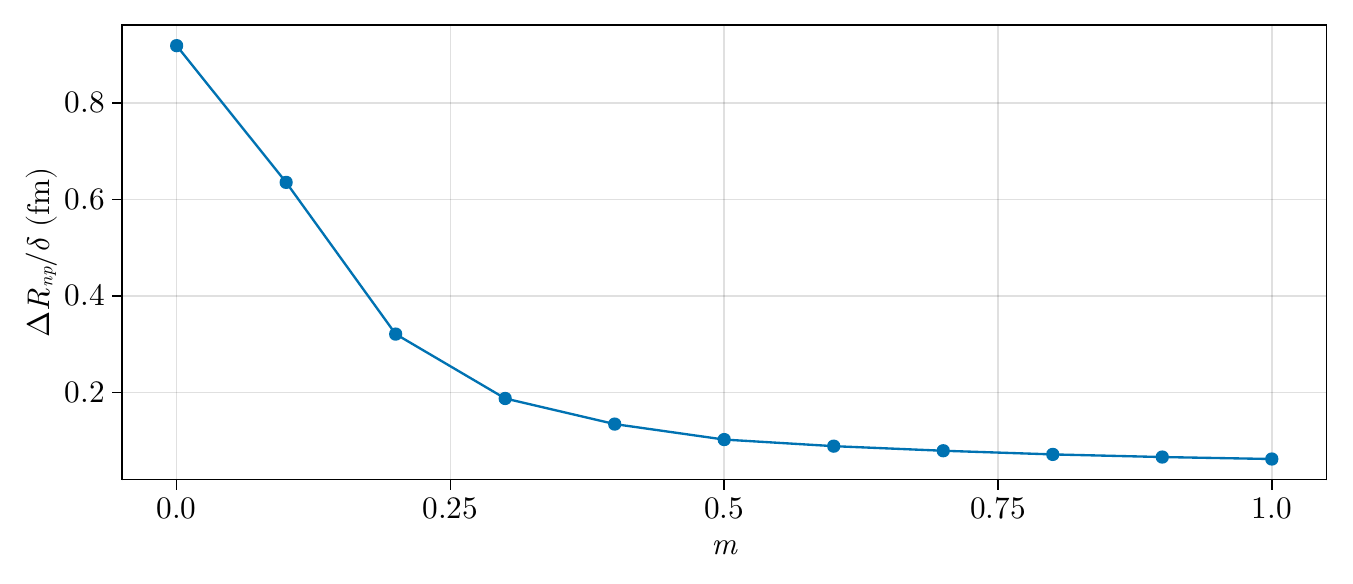}
    \caption{The average $\Delta R_{np}/\delta$ in $\fm$, as a function of the dimensionless pion mass parameter $m$.}
    \label{fig:as_fn_of_m}
\end{figure}

Our results depend on our calibration and hence on the pion mass $m$. To check this dependence, we calculate $\left(u^{(2)} / u r^2_{rms} - 1  \right)r_{rms} $ for a variety of $m$. To leading order, the important quantity is the constant part of this. We calculate it by taking the average over $B=108-265$. To calibrate the model, we must set the length scale of the Skyrme model which in turn sets the dimensionless pion mass, $m$. We will fix the length scale by comparing the baryon rms radius of the $B=108$ Skyrmion, $r_{108}$ to the radius of Tin-50 $r_{\text{Sn}}$ (Note: this calibration ensures the radii of all large nuclei are reasonably well described in a rigid rotor approximation). So in physical units, our result is
\begin{equation} \label{eq:physical_result}
\frac{\Delta R_{np}}{\delta} = \frac{r_{\rm Sn}r_{rms}}{r_{108}} \left(\frac{u^{(2)}}{u r^2_{rms}} - 1  \right)  \, \text{fm} \, .
\end{equation}
This is a function of the parameter $m$, which we can fit to match the linear relationship \eqref{Exprel}. We plot \eqref{eq:physical_result} as a function of $m$ in Fig. \ref{fig:as_fn_of_m}. The result is highly sensitive to $m$. The value of $m$ which gives the best fit is $m = 0.0$, giving
\begin{equation}
    \Delta R_{np} = 0.92 \delta
\end{equation}
This is to be compared to the result from experiments $\Delta R_{np} = 1.32 \delta$ \cite{Novario:2021low}, and the liquid drop model $\Delta R_{np} = 0.75 \delta$ \cite{pethick1996dependence}. 

Unfortunately, $m$ is not really a free parameter. It is related to the physical pion mass $m_\pi$. If we take $m_\pi$ equal to its physical value, $138$ MeV, we find that $m \approx 0.5$ and so $\Delta R_{np} = 0.15 \delta$. This is too small by an order of magnitude. That the pion mass, which should provide a small correction in chiral effective field theories, makes such a large difference is surprising.

\section{Relation to the Symmetry Energy}

The nuclear symmetry energy $S(n)$ is defined as the difference between the energy (per baryon number) of infinite, pure neutron matter and isospin symmetric nuclear matter at a given baryon density $n$. Although its dependence on the density has proven difficult to measure experimentally, it is usually parametrized as an expansion in powers of the baryon density around nuclear saturation $n_0\approx 0.16 \fm$,
\begin{equation}
    S_N(n_B)=S_0 +\frac{1}{3}L \epsilon +\frac{1}{18}K_{\rm sym}\epsilon^2 +\cdots
\end{equation}
with $\epsilon=(n-n_0)/n_0$, and 
\begin{equation}
    L=3 n_0\pdv{S_N}{n}\eval_{n=n_0}, \quad  K_{\rm sym}=9n_0^2\pdv[2]{S_N}{n}\eval_{n=n_0}
    \label{eq_symetobs}
\end{equation}
the slope and curvature of the symmetry energy at saturation, respectively. The symmetry energy at saturation is well constrained ($S_0\sim 30$ MeV) by nuclear experiments, but the values of the slope and higher order coefficients are still very uncertain (see e.g. the recent review \cite{Lattimer:2023rpe}).

The existence of a strong, model-independent correlation between the neutron skin thicknesses of neutron-rich nuclei, such as $^{48}{\rm Ca}$ and $^{208}{\rm Pb}$, and the symmetry energy slope parameter $L$ has been well established both from the experimental and theoretical ends \cite{Lattimer:2023rpe, Sammarruca:2023mxp}.
 Indeed, the symmetry energy slope can be understood as a pressure gradient acting on excess neutrons, which determines the formation and size of the neutron skin. Therefore, measurements of the neutron skin of nuclei can provide constraints on $L$, and hence on the nuclear matter equation of state.

In the Skyrme model, infinite nuclear matter can be described as a periodic crystal of half-skyrmions with cubic unit cells \cite{Klebanov:1985qi}. The symmetry energy of such crystals can be obtained in the rigid rotor approximation, which yields a simple expression in terms of the isospin moment of inertia of the unit cell $\Lambda$ \cite{Adam:2022aes}:
\begin{equation}
    S(n)=\frac{\hbar^2}{2\Lambda}.
\end{equation}
 
As with finite nuclei, the computation of the isospin inertia moment must be done numerically for each value of the unit cell length parameter $l$. Hence, there is not an analytic expression for $\Lambda(l)$. However, it was shown in \cite{Adam:2021gbm} that the crystal displays an almost \emph{perfect scaling} property with $l$. This property means that each of the terms in the energy functional \eqref{Energy_integral} scales with $l$ independently as $E_i \propto l^{-i+3}$, where $i$ is the number of spatial derivatives appearing in that particular term. 
This perfect scaling property is a characteristic of the field configuration and not only of its energy. Indeed, it is observed also for the isospin moment of inertia of the crystal \cite{Adam:2023cee}. 
Hence the (adimensional) energy $E$ and isospin moment of inertia $\Lambda$ of the cubic unit cell approximately satisfy the following expressions,
    \begin{align}
        E(l, m) &= K_2 l + \frac{K_4}{l} + m^2K_0 l^3,
        \label{e_PS}\\[2mm]
        \Lambda(l) &= \Lambda_2 l^3 + \Lambda_4 l,
    \end{align}
where the values of  $K_i$ and $\Lambda_i$ are ``universal'', i.e., they do not depend on either the parameters or $l$. The values of these universal scaling constants have been obtained in \cite{Adam:2023cee} and are given by
    \begin{align}
        K_0=0.017,\quad K_2 =&0.466,\quad K_4=9.617,\\
        \Lambda_2=0.038,&\quad \Lambda_4=1.393 .
    \end{align}
The above values are dimensionless, but were obtained using a different convention of Skyrme units. As we will be mainly interested in the behavior of $L$ with the pion mass parameter, and not on its specific value, we may as well use them.

Although the scaling is not perfect, in general the largest deviations from the true numerical values of energy start quite far from the minimum, at which the perfect scaling fit is most precise. Taking advantage of this almost perfect scaling property, we can write the symmetry energy slope at saturation as 
\begin{equation}
    L=3n_0\pdv{S}{n}\eval_{n_0}\equiv \frac{\hbar^2}{2}\frac{\Lambda'(l_0)}{\Lambda^2(l_0)}l_0
    \label{L(l0)}
\end{equation}
where $l_0$ is the unit cell length at saturation, i.e. at the minimum of $E(l,m)$ for any given $m$:

\begin{equation}
    \pdv{E(l,m)}{l}\eval_{l_0}=0\implies l_0=\sqrt{\frac{\sqrt{K_2^2+12m^2K_0K_4}-K_2}{6m^2 K_0}}.
    \label{l0(m)}
\end{equation}
Equations \eqref{L(l0)} and \eqref{l0(m)} above implicitly define a function $L(m)$, so we can study the correlation between the NST $\Delta R_{np}$ and the symmetry energy slope at saturation $L$ in the Skyrme model by comparing how they change with the pion mass parameter $m$.

\begin{figure}
    \centering
    \includegraphics[scale=0.6]{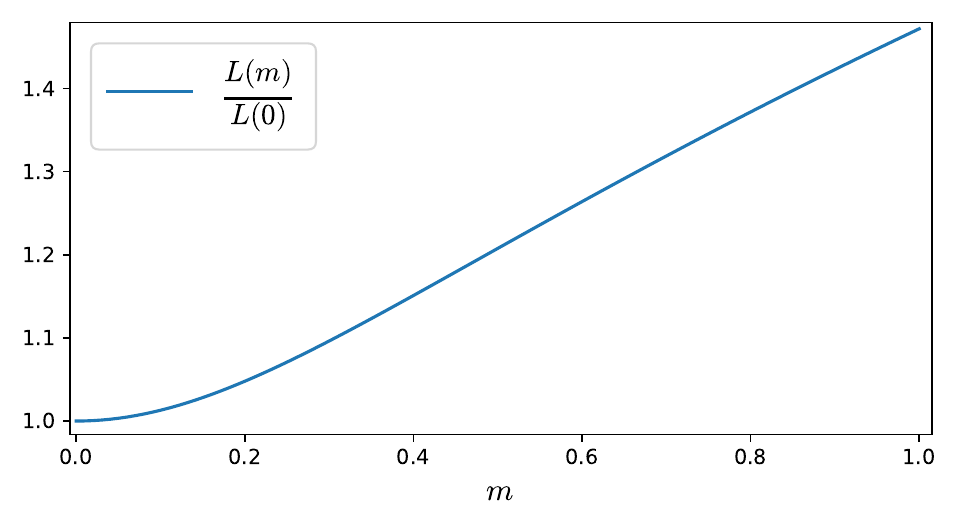}
    \caption{Symmetry energy slope as a function of the pion mass parameter in the Skyrme model.}
    \label{fig:L_m}
\end{figure}

In \cref{fig:L_m} we show the dependence of the symmetry energy slope $L$ with the pion mass parameter. Surprisingly, comparing with \cref{fig:as_fn_of_m}, we find that the symmetry energy slope and the neutron skin thickness are anti-correlated: while $L$ grows with increasing $m$, the NST decreases. This is precisely opposite to what was a priori expected from other nuclear models and constitutes a very unique feature of the Skyrme model.  

\section{Conclusions}

Overall, in the Skyrme model, the neutron skin thickness is proportional to the isospin asymmetry, with a constant of proportionality equal to the difference between matter and isospin charge radii (multiplied by the matter radius). This quantity has a weak dependence on baryon number, thus approximately reproduces the linear relationship seen in experimental data. Importantly, thanks to several approximations, we can derive almost everything analytically and find an expression for the neutron-skin thickness with a simple physical interpretation. This is a key advantage of the Skyrme model: using it, we can make analytical progress on difficult problems.

Using the physical pion mass, the constant of proportionality was much too small. This can be improved by decreasing the physical pion mass. There are several ways to interpret the result. The pion mass may be renormalized for large skyrmions, so a value different than the physical mass is reasonable to choose. Another possibility is that our calculation has only captured one contribution to the total neutron skin thickness, due to the simplicity of the original Skyrme model. A natural extension of the results presented here would be to include further interaction terms and/or degrees of freedom, which will affect the classical skyrmion solutions. For instance, vector mesons \cite{Fujiwara:1984pk}, or the sixth order term in $L_\mu$, first proposed in \cite{Jackson:1985yz} (see also \cite{Adkins:1983nw}), which can be seen as an effective point-like interaction that describes the repulsive exchange of omega vector mesons. The latter term has been shown to become relevant at sufficiently high densities (i.e. may be important to model large nuclei such as the ones we have been working with in this paper) and be crucial for an accurate description of the high-density equation of state of neutron stars \cite{Adam:2020yfv, Adam:2022cbs}. A final possibility is that we may have excluded an important theoretical consideration such as the Coulomb forces, which may alter the shape of large skyrmions and modify the final result.

\section*{Acknowledgements}

We thank Sven Bjarke Gudnason for calculating properties of the solutions from the smörgåsbord of skyrmions, and Christoph Adam and Miguel Huidobro for  discussions. Part of this work was completed during a visit of CH to Universidade de Santiago de Compostela; he thanks them for their hospitality. A. G. M. acknowledges financial support  from the PID2021-123703NB-C21 grant funded by MCIN/ AEI/10.13039/501100011033/ and by ERDF, ``A way of making Europe''; and the Basque Government grant (IT-1628-22). CH is supported by the Carl Trygger Foundation through the grant CTS 20:25. 


\section*{References}

\end{document}